\documentclass[notitlepage, aps, prb, 10pt, superscriptaddress, twocolumn]{revtex4-1}
\pdfoutput=1

\usepackage[utf8]{inputenc}

\usepackage{amsmath}
\usepackage{bm}
\usepackage{graphicx}
\usepackage[]{siunitx}
\usepackage{amsmath}
\usepackage{amsfonts}
\usepackage{amssymb}
\usepackage{amsthm}
\usepackage{mathtools} 
\usepackage{acronym}
\usepackage{color}

\usepackage[hyperindex,breaklinks]{hyperref}

\usepackage{comment}

\graphicspath{{./final_figures/}}
\newcommand{\figwidth}{3.325in}

\newcommand{\etal}{{\emph{et al.}}}

\newcommand{\RnB}[1]{ \left( #1 \right)}
\newcommand{\CrB}[1]{ \left\{ #1 \right\}}

\newcommand{\Ang}{\text{\normalfont\AA}}
\begin{document}

\title{Multi-scale modelling of current-induced switching in magnetic tunnel junctions using \textit{ab initio} spin transfer torques}
\author{Matthew O. A. Ellis}
\author{Maria Stamenova}
\author{Stefano Sanvito}
\affiliation{School of Physics, AMBER and CRANN Institute, Trinity College Dublin, Ireland}

\date{\today}
\begin{abstract}
    There exists a significant challenge in developing efficient magnetic tunnel junctions with low write currents 
    for non-volatile memory devices. With the aim of analysing potential materials for efficient current-operated 
    magnetic junctions we have developed a multi-scale methodology combining the \emph{ab initio} calculations 
    of spin-transfer torque with large-scale time-dependent simulations using atomistic spin dynamics. In this work 
    we introduce our multi-scale approach including a discussion on a number of possible mapping schemes the \emph{ab initio} 
    spin torques into the spin dynamics. We demonstrate this methodology on a prototype Co/MgO/Co/Cu tunnel junction 
    showing that the spin torques are primarily acting at the interface between the Co free layer and MgO. Using spin 
    dynamics we then calculate the reversal switching times for the free layer and the critical voltages and currents required 
    for such switching. Our work provides an efficient, accurate and versatile framework for designing novel current-operated
    magnetic devices, where all the materials details are take into account.

\end{abstract}
\maketitle

\section{Introduction}

Magnetic tunnel junctions (MTJs), composed of two epitaxially-grown ferromagnetic (FM) metal layers separated by an
insulating barrier (most often a few monolayers of MgO providing a dramatic spin filtering enhancement), constitute the principle
unit for a multitude of emerging technologies, in particular in Magnetic Random Access Memory (MRAM) and Spin Torque
Oscillators (STOs)\cite{Brataas2012,Ralph2008}. In both these cases the magnetisation dynamics of the free FM layer is driven by a
spin-polarised current. When the free layer magnetisation is misaligned with that of the polarising layer under current-carrying
conditions, the exchange interaction between the itinerant and localised electron spins results in a spin-transfer torque (STT), which
typically opposes the Gilbert damping torque and promotes switching\cite{Slonczewski1996}. For MRAM applications it is a
significant challenge to develop MTJs with a suitably low write current so as to ensure energy efficiency and to prolong device
lifetime\cite{Stamps2014}.

It is becoming increasingly more apparent that computational modelling can provide an initial analysis of the viability of
materials for efficient MTJs. However, only a few studies have been able to analyse a MTJ on multiple scales. Recent work has
focussed on developing more precise \emph{ab initio} models of spin-transfer torque\cite{Heiliger2008,Haney2008}, while
typical micromagnetic modelling employs Slonczewski's theory\cite{Berkov2008} and can sometimes ignore the fine atomic 
details of the system. Atomistic Spin Dynamics (ASD) has proved useful in modelling systems on a finer detail than micromagnetics
and has been developed to employ \emph{ab-inito} parameters to better describe the STT\cite{Chureemart2017}. Still
there remains a significant gap in our modelling ability, since to date no quantitative and materials specific transport method
has been combined with spin dynamics simulators. In practice this means that we are not capable of performing
current-induced spin dynamics simulations without making {\it a priori} assumptions on the nature and type of the STT.

In this work we attempt to bridge this gap and we present a multi-scale approach to modelling current-induced magnetisation
dynamics in magnetic devices using STT. At the microscopic scale a quantum transport method is employed to compute an
{\em ab initio} atom-resolved STT, which is then mapped onto the Landau-Lifshitz-Gilbert (LLG) equation of motion for atomistic
magnetic moments to perform the magnetisation dynamics\cite{Ellis2015,Evans2014}. The method is general and can be applied
to metallic and tunnelling junctions on the same footing, including nano-scaled objects such as point contacts or atoms on surfaces. 

Our paper is structured as follows; first we will introduce the computational scheme for calculating the {\em ab initio} STT and its 
mapping onto our atomistic spin model. We will then demonstrate this methodology on an example Co/MgO/Co/Cu MTJ stack.
We will discuss the bias, current and spatial dependence of the STT and how these features influence the magnetisation switching
of the free layer, both at zero and finite temperature.

\section{Methods}
Our multiscale methodology is built upon using an {\em ab initio} method at the microscale for the electron transport and an
atomistic scale spin model to simulate the dynamics. In particular we utilise the {\sc Smeagol}\cite{Rocha2006,Rungger2009}
code to model ballistic electron transport through the MTJ under a finite bias voltage. {\sc Smeagol} is an implementation of the
Keldysh non-equilibrium Green function (NEGF) approach to the steady-state open-boundary problem within the framework of
Density Functional Theory (DFT), as implemented in the {\sc Siesta} code, which provides an efficient order-$N$ scalling core DFT algorithm\cite{Soler2002}. Within this formalism the MTJ is modelled
as a central scattering region (SR) connected to two semi-infinite periodic leads. As the electronic properties of the latter can be
determined independently from those of the junction their action on the scattering region can be described in terms of suitably 
chosen self-energy operators acting at the SR boundaries. This effectively reduces the original electronic structure problem for an infinite 
non-periodic system to an energy dependent problem for a finite atomic construct. The bias voltage, $V$, is applied as a shift to the chemical 
potentials of either lead by $\pm V/2$ and the non-equilibrium charge density of the SR can be determined self-consistently
from the associated non-equilibrium Keldysh Green's function.

For our calculation of the spin-transfer torque we follow the approach proposed by Haney \etal\cite{Haney2007}. The 
out-of-equilibrium spin density, $\boldsymbol{\sigma}^{V}$, is assumed to be separable into an equilibrium spin density, 
$\boldsymbol{\sigma}^{0}$, and a transport correction, $\boldsymbol{\sigma}^\mathrm{tr} $, where such correction is much 
smaller in magnitude than the equilibrium part. A transverse spin transport contribution arises from the non-collinearity 
in the open-boundary system giving rise to a STT in the free layer. Further details of our method are given in 
Ref.~[\onlinecite{Stamenova2016}]. Here we adopt the magnetic moment version (as opposed to working with spin variables) of the atom-resolved STT, in which the STT acting on the $a$-th atom is written as
\begin{equation}
\mathbf{T}_a = \frac{\mu_\mathrm{B}}{2} \sum_{i \in a} \sum_j \boldsymbol{\Delta}_{ij} \times \boldsymbol{\sigma}_{ji}^\mathrm{tr}\:,
\label{eq:torque}
\end{equation}
where $\boldsymbol{\Delta}_{ij}$ are the matrix elements of the exchange-correlation field written over the localised atomic 
basis orbitals of {\sc Siesta} and $\mu_\mathrm{B}$ is the Bohr magneton. Note that while the first summation is restricted to 
orbitals that belong to the atomic site $a$ (the atom for which the torque is calculated), the second one spans over all the 
orbitals in the SR. The transport spin is calculated from the difference between the equilibrium ($V=0$) and the non-equilibrium 
($V\ne0$) density matrices, $\rho_{ij}^{V}$, as
\begin{equation}
\boldsymbol{\sigma}^\mathrm{tr} = \text{Tr} [(\rho^V - \rho^0) \boldsymbol{\sigma} ]\:,
\end{equation}
with $\boldsymbol{\sigma}$ being the vector of Pauli matrices.

The {\em ab initio} side of our multiscale approach is then completed with the evaluation of the dataset 
$\CrB{\mathbf{T}_a \RnB{V,\theta}}$ of atom-resolved STTs as a function of the bias voltage, $V$, and the angle, $\theta$, 
between the fixed and the free layer magnetisations. It should be noted here that the use of a single angular parameter assumes that there 
is no non-collinearity within the free layer. In some cases, when the self-consistent calculation of the density matrix across a 
range of finite-bias grid points is too involved computationally, we also utilise the linear response quantity, namely the 
spin-transfer torkance (STTk), $\boldsymbol{\tau}_a$, that is defined as
\begin{equation}
\boldsymbol{\tau}_a \equiv \frac{\partial \mathbf{T}_a}{\partial V} = \frac{1}{2} \sum_{i \in a}\sum_j \boldsymbol{\Delta}_{ij} \times \mathrm{Tr} \left[ \frac{\partial \rho_{ji}(V)}{\partial V} \boldsymbol{\sigma} \right]_{V=0} \:.
 \label{eq:torkance}
\end{equation}

Once the spin-transfer torques, $\CrB{\mathbf{T}_a\RnB{V,\theta}}$, for the given junction are obtained we can then proceed to
computing the current-induced magnetisation dynamics using an atomistic spin model. ASD is a semi-classical 
model typically using a Heisenberg spin Hamiltonian to describe a system of constant spin magnetic moments. These magnetic moments are localised at 
atomic sites and their dynamics is calculated from evolving discretised LLG-like equations of motion. The LLG equations for atomic spins with 
additional STTs are often referred to as LLG-Slonczewski equations, whose atomistic form reads
\begin{equation}
\frac{\partial \mathbf{S}_i}{\partial t} = -\gamma \mathbf{S}_i \times  \mathbf{H}_i +
\lambda \mathbf{S}_i \times \frac{\partial \mathbf{S}_i}{\partial t} + \frac{1}{\mu_i}\mathbf{T}_i(V, \left\{\mathbf{S}_i\right\})\:,
\label{eq:LLG}
\end{equation}
where $\mathbf{S}_i=\boldsymbol{\mu}_i/\mu_i$ is a unit vector in the direction of the spin magnetic moment of atom $i$ of
magnitude $|\boldsymbol{\mu}_i|=\mu_i$. Since the \emph{ab initio} torque in Eq.~(\ref{eq:torque}) is derived as the rate 
of change of the spin angular momentum it is necessary to normalise the torque to the unit vector used in the ASD.
In Eq.~(\ref{eq:LLG}) $\lambda$ is the atomistic damping parameter that corresponds to the Gilbert damping parameter at the 
microscopic scale and
\begin{equation}
\mathbf{H}_i (t) = - \frac{1}{\mu_{i}} \frac{\partial \mathcal{H}}{ \partial \mathbf{S}_i}  + \boldsymbol{\xi}_i (t) \label{eq:eff_field}
\end{equation}
is the effective magnetic field acting on spin $i$. The system is kept at a finite temperature through a stochastic time-dependent thermal field, 
$\boldsymbol{\xi}_i(t)$. In the white noise limit this is represented as a Gaussian random number with the following moments
\begin{align}
\langle \xi_{ia}(t) \rangle  &= 0\:, \\
\langle \xi_{ia}(t) \xi_{jb}(t') \rangle  &= \frac{2 \lambda k_B T}{\mu_s \gamma} \delta_{ij} \delta_{ab} \delta(t-t')\:,
\end{align}
where $i,j$ label the different atoms, $a,b = x,y,z$ are the Cartesian components and $t,t'$ is the time. In order to model the dynamics of
an MTJ free layer we limit the Hamiltonian to contain only the Heisenberg exchange and a uniaxial anisotropy term as follows
\begin{equation}
\mathcal{H} = - \sum_{ij} J_{ij} \mathbf{S}_i \cdot \mathbf{S}_j - \sum_i k_i (\hat{\mathbf{e}}_\text{ani} \cdot \mathbf{S}_i )^2\:,
\end{equation}
where $J_{ij}$ is the isotropic exchange constant and $k_i$ is the uniaxial anisotropy constant for spin $i$ along the axis
$\hat{\mathbf{e}}_\text{ani}$. In general one must also consider the demagnetising field acting on the free layer and its contribution 
to the anisotropy. In the following we consider the intrinsic anisotropy to be out-of-plane ($\hat{\mathbf{e}}_\text{ani} = \hat{\mathbf{z}}$) 
and since our free layer is ultra-thin the demagnetising field can be represented as that of an infinite thin platelet. Therefore, instead 
of calculating the demagnetising field directly, which can be costly since it involves adding long-range dipolar interaction to the spin
Hamiltonian, we incorporate it into the uniaxial field such that $k_i = k_u - \mu_0 (M_s V_\text{a})^2 /2$. Here $k_u$ is the intrinsic 
uniaxial anisotropy constant, $\mu_0$ is the permeability of free space, $M_s$ is the saturation magnetisation and $V_\text{a}$ is 
the atomic volume.

The next step is to map the two-parameter discretised {\em ab initio} $\CrB{\mathbf{T}_a\RnB{V,\theta}}$ dataset onto the STT term of
Eq.~(\ref{eq:LLG}) which is, in general, a continuous function of the angular coordinates of the whole set of spins $\left\{\mathbf{S}_i\right\}$. Such mapping can be performed in several manners and here we have implemented three different strategies.
The first is a full 2D interpolation of the dataset, i.e. for each atom $i$ in layer $l_i$ an interpolated STT value is obtained for the
specified voltage $V$ and the instantaneous angle $\theta = \text{acos}(\mathbf{S}_i \cdot \hat{\mathbf{P}} )$ between the local
spin $\mathbf{S}_i$ and the direction of the fixed layer magnetisation $\hat{\mathbf{P}}$. In order to simplify the calculation during
the simulations a linear interpolation is performed along $V$, while a cubic spline is used for $\theta$, since the dynamics is more
sensitive to the angular variation and only a limited set of angles are calculated at finite voltage.

Our second mapping uses the angular dependence of the STT derived by Slonczewski~\cite{Slonczewski2005}. In this way we
avoid calculating the angular dependence of the STT at each voltage from first principles. We note that Slonczewski's model is ideally valid for sufficiently wide momentum-filtering barriers~\cite{Slonczewski2005}. The torque magnitude, however, is taken from the \emph{ab initio} calculations, i.e. the bias dependence
of the torque is still from first principles, namely it is interpolated out of the \emph{ab initio} dataset. This semi-functional mapping
is given as
\begin{align}
\mathbf{T}_i(V,\mathbf{S}_i)  = &
 T_{||}(V, l_i) \mathbf{S}_i \times \mathbf{S}_i \times \hat{\mathbf{P}}  + 
 T_{\perp}(V, l_i) \mathbf{S}_i \times \hat{\mathbf{P}}\:, \label{eq:semifunc}
\end{align}
where $T_{||}$ and $T_\perp$ are the parallel and perpendicular torque magnitudes,
which can be extracted at $\theta = 90^\circ$.

Our final mapping utilises the torkance instead of the finite voltage torques. In this manner a finite voltage is simulated
by assuming a linear voltage dependence and by scaling the torkance to the desired $V$ as follows
\begin{equation}
 \mathbf{T}_i(V,\mathbf{S}_i, l_i) =  V \left. \frac{\partial \mathbf{T}(\theta, l_i)}{\partial V} \right\rvert_{V=0}\:.
\end{equation}

We discuss the applicability of this linear dependence in the case of a Co/MgO-based MTJ in the following section. The angular
dependence can again be interpolated using cubic splines, but it is also possible to also use the Slonczewski form given in
Eq.~(\ref{eq:semifunc}).

Although the STTs are extracted from ballistic transport at a constant bias voltage, we have developed a numerical scheme to 
utilise the {\it{ab-initio}}-calculated $I$-$V$ characteristics, which allows us to simulate the atomistic spin dynamics also under 
constant-current conditions. As we will show in the next section the conductance of a CoMgO-based MTJ is found to follow the 
equation
\begin{equation}
  g(\theta, V) = \frac{V}{J(V,\theta)} = A(V) + B(V) \cos(\theta)\:.
  \label{eq:gtheta}
\end{equation}
Our model can then compute the current as it changes with the free layer angle and apply the torque appropriately for the given 
current and voltage. This is directly reflected in the pre-factor of the Slonczewski's STT equations\cite{Slonczewski2007}.

\section{Results}

\subsection{{\em Ab Initio} STT in a Co-MgO MTJ}

\begin{figure}
  \centering
  \includegraphics[width=\figwidth]{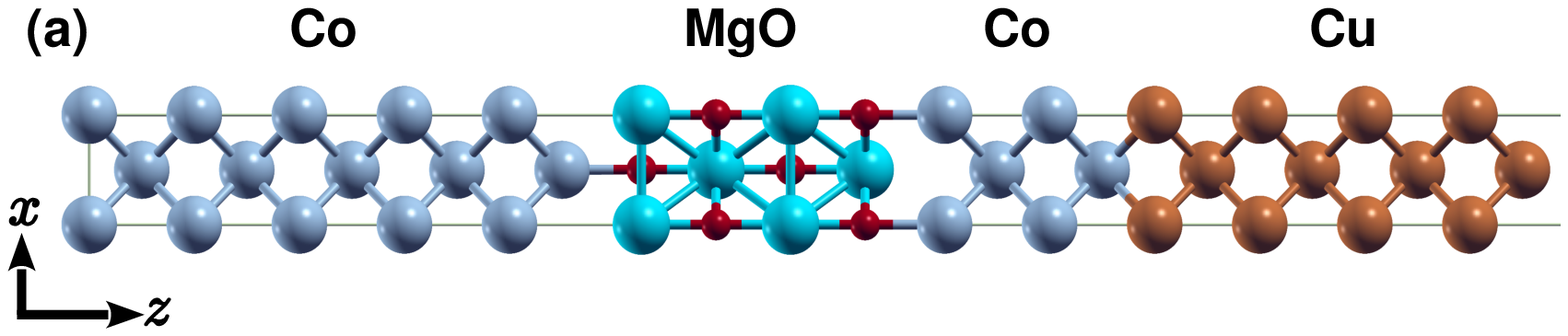}
  \includegraphics[width=\figwidth]{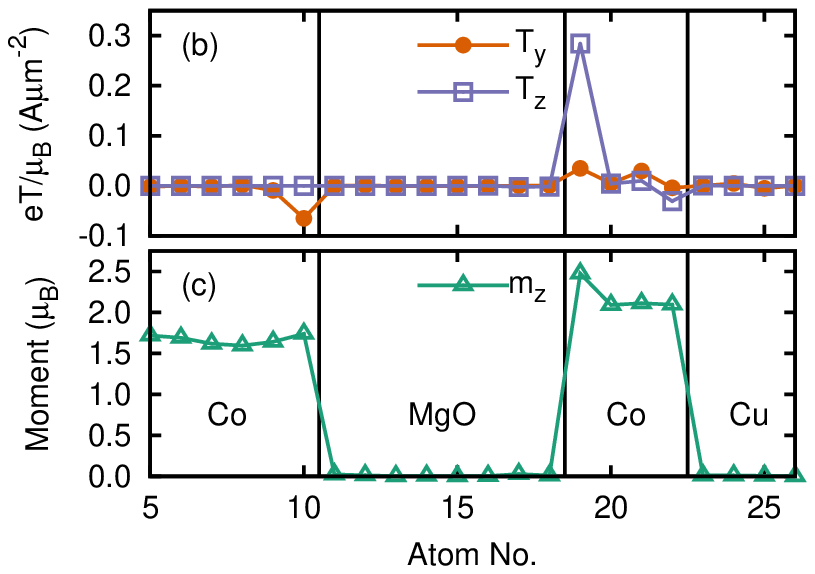}
  \caption{The Co/MgO MTJ stack studied in this work. Panel (a) shows a schematic of the scattering region for the
  {\sc Smeagol} calculation, while panels (b) and (c) present the atomic resolved \emph{ab initio} STT at \SI{1}{V} and 
  the atomic spin moments profiles, respectively. In (b) and (c) the first 4 Co and last 4 Cu atoms are omitted since in 
  the calculations these are replaced with the semi-infinite leads.}
  \label{fig:STT_layer}
\end{figure}

Our computational strategy is now tested for a CoFeB-MgO based MTJ, which is probably the most studied magnetic device today.
In order to model such system we simplify the structure to only comprise of Co atoms in a Co/MgO(4)/Co(4)/Cu stack, where
the numbers indicate the number of atomic planes in each layer. Note that the outermost layers are the semi-infinite leads as
visualised in Fig.~\ref{fig:STT_layer}(a). In our generic Co-based MTJ both leads share a {\it bcc} lattice with a lattice parameter of
2.857~\Ang. This is the lattice constant of Fe and the idea is to mimic the highly spin-polarised conventional CoFeB lead. 

Our DFT calculations are based on the local spin-density approximation with the Ceperley-Alder parameterisation of the 
exchange-correlation functional as implemented in the {\sc Siesta} code. A double-zeta numerical atomic basis set is used for 
all atomic species with additional polarisation for $s$-orbitals of the transition metal atoms. A Monkhorst-Pack Brillouin zone 
sampling is used, based on a 20$\times$20 real-space grid.

The magnetic moments of each layer are shown in Fig.~\ref{fig:STT_layer}(c). As expected there is no magnetisation in MgO and 
Cu, while the Co fixed layer shows moments close to the bulk value of $\mu_\text{Co}= 1.72 \mu_\mathrm{B}$. Since the free layer 
is ultra-thin the moments are larger than in the bulk with a peak at the MgO interface. From the layer resolved calculations we observe 
that the STT is strongly peaked at the MgO interface, as shown in Fig.~\ref{fig:STT_layer}(b) at 1~V. Following the sharp decay of the 
STT inside the Co layer, there is a characteristic higher STT value also at the other interface with the Cu lead but with an opposite sign.
\begin{figure}
  \centering
  \includegraphics[width=\figwidth]{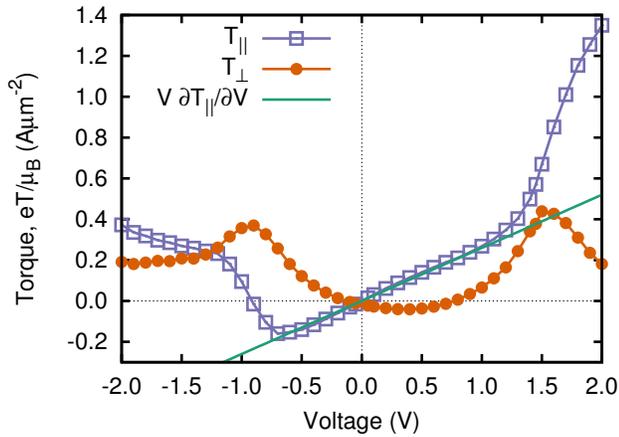}
  \caption{The voltage dependence of the in-plane (open squares) and out-of-plane (filled circle) torque, and the in-plane torkance
  (solid line). The in-plane torque shows a linear behaviour up to approximately \SI{1.4}{V}. Within this range the torkance is a good
  approximation of the finite bias torque. The out-of-plane torque shows a quadratic-like behaviour, for which the zero-bias torkance
  is not sufficient to describe. }
  \label{fig:STT}
\end{figure}

\begin{figure}
  \centering
  \includegraphics[width=\figwidth]{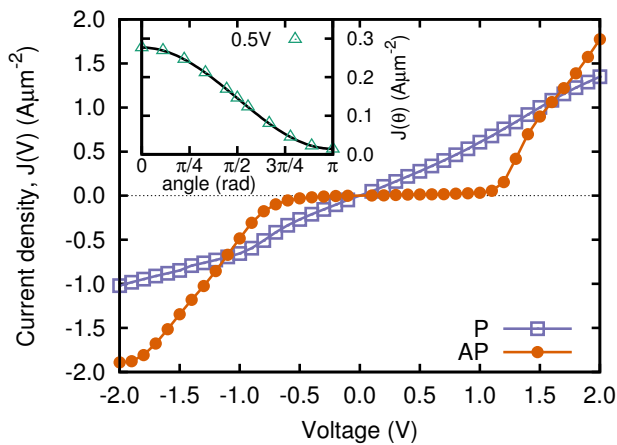}
  \caption{The resulting current density for an applied bias voltage in the Co/MgO/Co/Cu MTJ. The solid circles show the current
  density in the anti-parallel configuration, while the open squares show the parallel configuration. Up to approximately 1~V there
  is a significant TMR but above this value more current flows in the anti-parallel state and the TMR drops. The inset shows the
  angular dependence of the current density at \SI{0.5}{V}, which is fitted by using $J(\theta) = A + B \cos(\theta)$ (solid line). }
  \label{fig:IV}
\end{figure}

Figure~\ref{fig:STT} shows the total STT acting on the free layer in the Co-MgO MTJ as function of the applied bias voltage for a
fixed misalignment of the free layer magnetisation of 90$^\circ$. The asymmetry of the torque with bias arises from the asymmetry 
of the stack, namely the free layer contains only 4 atomic planes, while the fixed layer in our MTJ is semi-infinite. In both cases, 
however, there is approximately a linear and a quadratic relationship with voltage for the out-of-plane and in-plane torques, respectively. 
The slope of the in-plane STT around zero matches well our zero bias torkance from Eq.~(\ref{eq:torkance}), therefore the latter 
approximation offers a reasonable quantitative measure for the in-plane STT at low bias.

Figure~\ref{fig:IV} shows the current-voltage characteristics for our MTJ stack in both the parallel (P) and anti-parallel (AP) configuration.
The sharp increase of the in-plane STT above 1.4~V in Fig.~\ref{fig:STT} is due to the increase of the conductivity in the anti-parallel
configuration. This is in turn due to the fact that the $\Delta_1$ symmetry band for the minority spin carriers is approximately aligned 
to the $\Delta_1$ majority one at that bias voltage~\cite{Xie2016}. Intriguingly whilst this leads to a lower TMR at high voltages the 
increased electron flow appears to result in a larger in-plane torque and in a reduction of the out-of-plane one, as can be seen in 
Fig.~\ref{fig:STT}. The inset to Fig.~\ref{fig:IV} shows the variation in the current density due to the misalignment angle of the FM 
layers. For nearly all the voltages simulated the current can be modelled by using Eq.~(\ref{eq:gtheta}).

\subsection{Switching dynamics at zero temperature}

We now move our attention to the switching dynamics based on the {\em ab initio} torques computed in the previous section. 
In order to construct the spin model we require values for the exchange constants, uniaxial anisotropy, atomistic Gilbert damping 
and magnetic moments. For the exchange we use the tabulated bulk value\cite{Evans2014} for {\it bcc} Fe, namely $J_{ij} = \SI{7.05e-21}{J}$, 
which is assumed here to be similar to that of {\it bcc} Co, whilst the magnetic moments are taken directly from the {\sc Smeagol} 
calculations. In order to explore a wide range of current induced switching we vary the anisotropy between \SI{0.001}{meV} and 
\SI{0.5}{meV} which, as discussed earlier, accounts for both intrinsic anisotropy and demagnetising field. First principles calculations by Hallal \etal\cite{Hallal2013} on Fe/MgO thin films found that the anisotropy is $k_u \approx \SI{0.275}{meV}$ per atom for a layer thickness similar to ours. For comparison the switching field at $k=\SI{0.1}{meV}$ is $H_k \approx \SI{1.7}{T}$, while to achieve a thermal stability of
$K V / k_B T_\text{room} =60$  an area of $(\SI{36}{nm})^2$ is required. The Gilbert damping in thin films has been observed
to vary with the layer thickness and the presence of capping layers can enhance the damping through spin pumping effects. 
Experimental measurements for a Ta/CoFeB/MgO stack show damping parameters of the order $\lambda=0.01$ for ultra-thin FM 
layers\cite{Iihama2015} and so here we vary the damping from 0.01 to 0.1. The magnetisation dynamics is computed by numerically 
solving Eq.~(\ref{eq:LLG}) using the Stochastic Heun scheme\cite{Evans2014} with a time-step of \SI{0.1}{fs}. This has been tested for 
stability in equilibrium.

We start by investigating the voltage required to observe switching in the MTJ free layer without explicit thermal effects. The lack of 
thermal effects allows us to simulate the switching with only the basic unit cell and periodic boundary conditions in the lateral directions. 
In order to measure the switching we calculate the time that is required for $m_z$ to pass the $m_z=0$ plane. We model the dynamics of
each MTJ by initiating the simulation with a small deviation of the free-layer magnetisation from the $-\mathbf{\hat{z}}$ axis at different 
applied bias voltages.

\begin{figure}
	\centering
    \includegraphics[width=\figwidth]{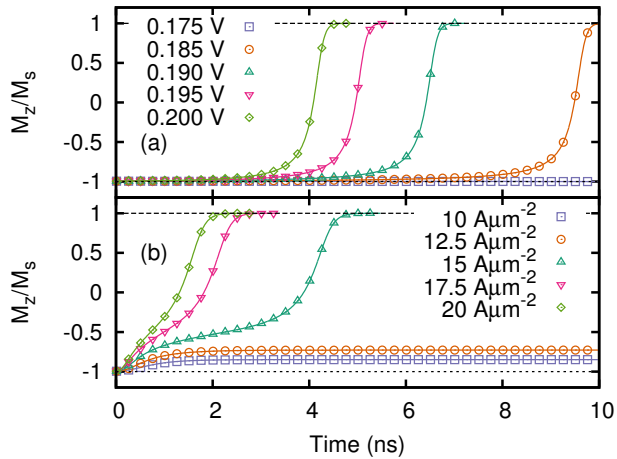}
    \caption{Magnetisation switching for a junction kept at (a) constant voltage and (b) constant current for $\lambda = 0.01$ and
    $k_i = \SI{0.1}{meV}$. At constant voltage the switching is uniform above the critical voltage, while at constant current the torque
    has an additional angular dependence given by the variation of the conductivity (hence the voltage at constant current) with angle.}
	\label{fig:Mzvtime}
\end{figure}

The magnetisation switching curves are shown in Fig.~\ref{fig:Mzvtime} for (a) constant voltage and (b) constant current
by using an anisotropy of $k = \SI{0.1}{meV}$ and a damping parameter of $\lambda = 0.01$. When the junction is kept at a 
constant voltage the switching is uniform and stable. In practice the magnetisation of the free layer remains anti-parallel to 
that of the pinned one for a long time and then switches fast. This is expected since the torque increases as the two 
magnetisation vectors become non-collinear and it is maximised for $\theta=90^\circ$. Furthermore, it is observed that increasing 
the voltage systematically shifts the transition to lower times. 

In contrast, at a constant current the torque can initially overcome the anisotropy but, as the misalignment angle between the 
fixed and the free layer decreases, the resistance of the junction also decreases. This causes the voltage required to maintain 
the desired current to be reduced, and as a consequence also the torque is reduced. The reduction of the torque as the 
magnetisation vectors become non-collinear to each other has to be contrasted with an increase of the anisotropy, leading to 
a stable precessional state where a fine balance of the torques is achieved. As the current is increased further the angle of this 
stable point becomes larger until it reaches the maximum of the anisotropy torque at about $45^\circ$. Then the full reversal 
occurs. Further increasing the current reduces the reversal time and also the transition width.

\begin{figure}
	\centering
  	\includegraphics[width=\figwidth]{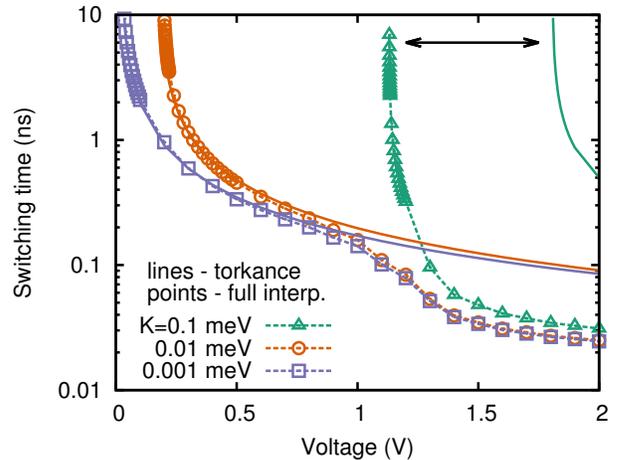}
  	\caption{The switching time for a Co free layer as a function of bias voltage for three values of the anisotropy and a damping 
	coefficient of $\lambda = 0.1$. The open points are for calculations performed with the full interpolation, while the solid lines 
	are for the torkance method and the dotted ones are a guide to the eye. The arrow indicates the difference between the torkance 
	and full interpolation methods for the K=\SI{0.1}{meV} case.}
    \label{fig:switching_times}
\end{figure}

Figure~\ref{fig:switching_times} shows the measured switching time against the voltage calculated with the different mapping 
strategies for three values of the anisotropy. We find that there is no signficant difference between the full and semi interpolation 
methods since the angular dependence of the \emph{ab initio} STT agrees well with the Slonczewski form. As such only the full 
interpolation results are compared to the torkance-based ones. For each anisotropy there is no switching below a critical voltage 
and a sharp decay of the switching time above it. Since there is a large increase in the torque above approximately 1.4~V 
(see Fig.~\ref{fig:STT}), the switching time shows a consistent drop at this point. For an anisotropy of \SI{0.1}{meV} (green triangles 
and line) the critical voltage lies close to this increased torque and we find that there is a large difference between the calculations 
using the finite-voltage torques and those obtained at zero-voltage with the torkance method.

\begin{figure}
  \centering
  \includegraphics[width=\figwidth]{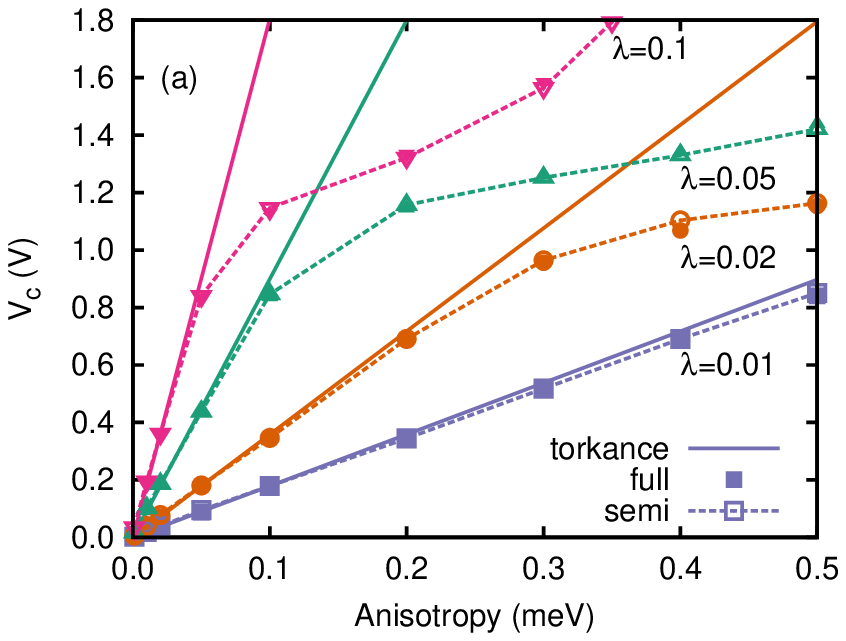}
  \includegraphics[width=\figwidth]{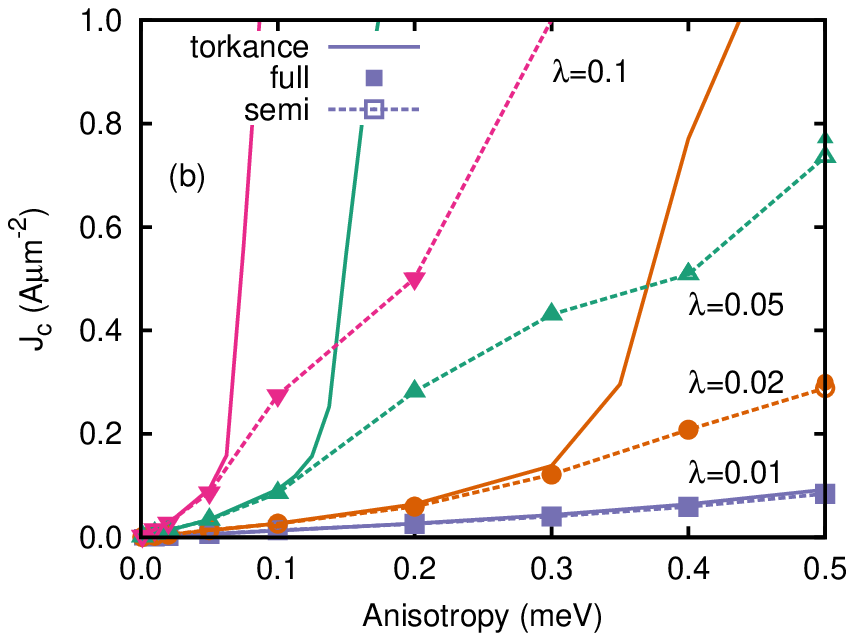}
  \caption{The critical (a) voltage and (b) current required to switch the free layer for a given anisotropy and damping at $T=0$~K.
  Three alternative methods for interpolating the STT are shown for each case; torkance (solid lines), full interpolation (filled circles)
  and semi-functional (open circles). The dotted lines are a guide to the eye.}
  \label{fig:Vc_T0K}
\end{figure}

The critical voltages and currents for a range anisotropy strengths and damping coefficients are shown in Fig.~\ref{fig:Vc_T0K}.
The three interpolation methods discussed earlier are shown as solid lines for the torkance, filled points for full interpolation and
open points for the semi-functional method. Our results show that there is no significant difference between the semi-functional
and the full interpolation method over the range simulated here. For the full interpolation method the loss of numerical accuracy
may arise in some instances due to the poor interpolation at $\theta$ close to end points, $0$ and $\pi$, if too few data points are 
available where curvature is high. Such numerical errors lead to longitudinal torques, which effectively (due to the constrained 
spin length in the ASD) reduce the net torque.

The non-linear behaviour of the critical voltage shown in Fig.~\ref{fig:Vc_T0K}(a) arises simply because of the calculated voltage
dependence of the in-plane torque, while in (b) there is an additional effect arising from the voltage dependence of the current. 
At a lower damping the torkance matches the other methods for a wider range of anisotropies. This is due to the fact that the critical 
voltage is related to the product of the damping and the anisotropy. When the critical voltage is below approximately \SI{1}{V}, then
the torque is in the linear regime, hence, we find the torkance to agree well with the finite-voltage-calculated torque (see Fig.~\ref{fig:STT}). 
In high anisotropy systems, where a large switching voltage may be required, an accurate knowledge of the STT voltage dependence 
becomes important.

\subsection{Switching dynamics at finite temperature}

Finally, we consider the switching process at finite temperature. Now our simulation cell needs to be largely increased in order to
account for the temperature-induced non-collinearity. In this case we simulate a 32$\times$32$\times$4 spin slab corresponding to a 
lateral dimension of $\SI{9.2}{nm}$ and still apply periodic boundary conditions in the lateral directions. Ideally one should consider 
thermal effects on the current and the STT as well, but here we only consider thermal effects in the ASD through the stochastic noise 
term introduced into the effective field in equation (\ref{eq:eff_field}). The non-collinearity now requires a further decision when mapping 
the STT to the ASD. The {\em ab initio} calculation of the torque is for a fully collinear free layer but non-collinearity in ASD is 
required to achieve a thermal spin distribution. One can then decide to use the angle of the total magnetisation or that of each individual 
spin in order to determine the torque. The effects of this choice will be discussed in what follows. Note that, in principle, one can still
calculate the torques from {\it ab initio} for a non-collinear situation. In fact, one can even calculate the torques at each time step in the 
ASD, as it is done for instance for the forces in {\it ab initio} molecular dynamics. This is, however, not practical here since the transport 
calculations, in particular at finite bias, are much more demanding than the ASD ones. 
\begin{figure}
\centering
\includegraphics[width=\figwidth]{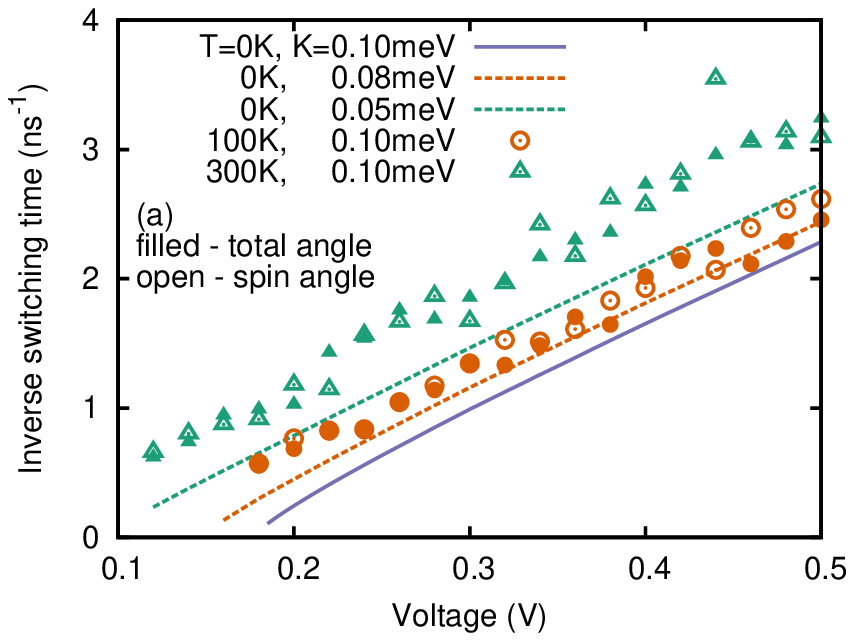}
\includegraphics[width=\figwidth]{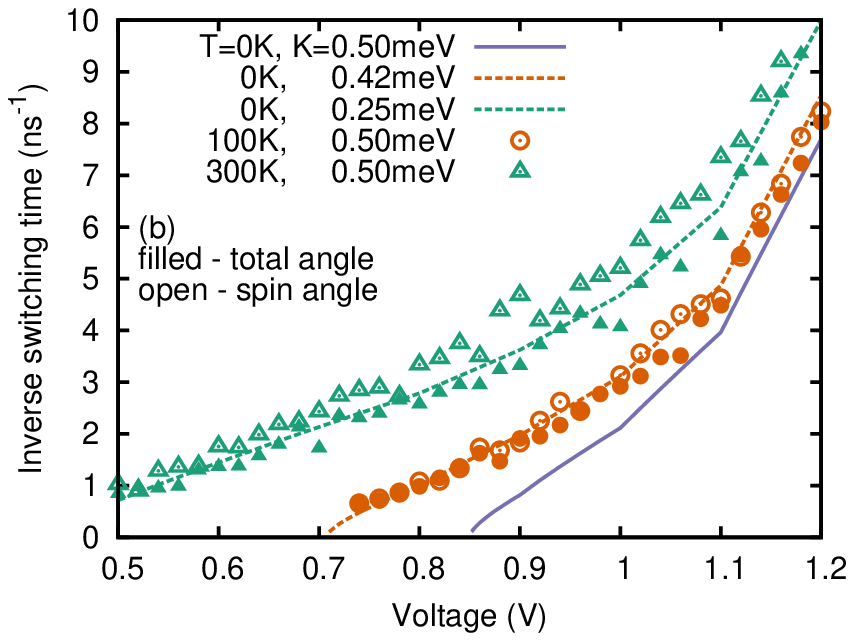}
\caption{Inverse switching time with (a) $k=0.1$~{meV} and (b) \SI{0.5}{meV} at $T=0$~{K} (solid blue line), \SI{100}~{K} (orange circles) 
and \SI{300}~{K} (green triangles). Filled and open symbols represent simulations run by using the angle calculated for the total magnetisation 
or for each individual spin, respectively. The dashed lines indicate the inverse reversal time at $T=0$~{K} using a scaled anisotropy constant.}
\label{fig:Temp_invtime}
\end{figure}

Figure~\ref{fig:Temp_invtime} shows the inverse average switching time at different temperatures for (a) $k=0.1$~{meV} and 
(b) 0.5~{meV}. The filled symbols show results obtained by using the angle of the total magnetisation to calculate the STT, 
while the open ones use the individual spin angle. From the figure we observe that results obtained with the different angle 
methods are almost indistinguishable from each other except in (b) at \SI{300}{K}. Here the switching time is averaged over 
24 independent simulations since it is a stochastic process. This may lead to an equivalence of methods, since whilst these 
are fundamentally different the average switching time may be similar.

Different anisotropies present us two different situations. In Fig.~\ref{fig:Temp_invtime}(a) the inverse relaxation time is linear
with the voltage since the critical voltage is within the linear regime, while in Fig.~\ref{fig:Temp_invtime}(b) it is non-linear. In general, however, for both anisotropy values  
increasing the temperature reduces the switching time and also the critical voltage. Within a micro-magnetic picture this behaviour 
is reproduced by introducing temperature dependent parameters, namely the anisotropy, the damping and the magnetic moment. 
These reduced parameters then lead to a reduction in the critical switching voltage. Callen-Callen theory\cite{Callen1963} predicts that at finite temperature the macroscopic 
uniaxial anisotropy constant, $K_u$, scales as $K_u(T)/K_u(0) = [M(T)/M(0)]^3$. From our simulations we find that at \SI{100}{K} 
and \SI{300}{K} the average magnetisation is approximately 0.94 and 0.80 respectively. This returns us expected anisotropy constants 
of $K_u(100) \approx 0.83 K_u(0)$ and $K_u(300) \approx 0.51 K_u(0)$. The dashed lines in figure~\ref{fig:Temp_invtime}, 
therefore show the inverse switching time at \SI{0}{K} obtained by using these scaled anisotropy values. As we can see in 
panel (b) the zero-temperature dynamics computed using these scaled constants agree well with the average switching time  
obtained at finite temperature despite the lack of thermal fluctuations. The same is not true for the lower anisotropy case of 
Fig.~\ref{fig:Temp_invtime}(a). Here there is agreement only at higher voltages for \SI{100}{K}, while at \SI{300}{K} the 
zero-temperature switching times at the re-scaled anisotropies are constantly longer than those obtained with the 
finite-temperature dynamics. This has to be attributed to the actual thermal fluctuations, which are more pronounced for a lower 
anisotropy and cause the switching to occur faster.


\section{Conclusion}

To summarise, we have developed a multi-scale modelling methodology combining {\em ab initio} calculations of the spin-transfer torque
and large-scale finite-temperature spin dynamics simulations. Using the {\sc Smeagol} code, both the STT and the STTk have been computed
for various applied voltages and angles of misalignment between the fixed and free magnetic layer in a nano-scopic junction. This is then mapped onto an atomistic spin
dynamics model, which is used to calculate the switching times with and without thermal effects. We apply this methodology to a prototype
MTJ based on Co/MgO, where we find that the STT is strongly localized on the Co atoms at the MgO interface and that the STT is linear
at low voltages. In contrast above 1.4~V there is a sharp increase in the total current driven by the minority spin component. Such current
density increase leads to a sharp enhancement of the in-plane torque and in a reduction of the out-of-plane one. 

The {\it ab initio} calculated torques are then
mapped onto the spin dynamics with different mapping types being analysed. A full interpolation of the \emph{ab initio} data set is preferred
but using the Slonczewski angular form together with the {\it ab initio} voltage dependence extracted at a fixed angle performs equally well
over a wide range of parameters. Due to the linear nature of the STT at low bias the 0~V linear response (torkance) is a suitable replacement. At finite temperature the picture described above does not change drastically, except for the
fact that the thermal fluctuations reduce the critical voltage required for switching. Thus, we have demonstrated that our multi-scale
construction offers a viable approach for the characterisation and ultimately design of current-driven magnetic devices.

\section{Acknowledgements}
This work has been supported by the Science Foundation Ireland Principal Investigator award (grant no. 14/IA/2624 and 16/US-C2C/3287).
We gratefully acknowledge the Irish Center for High-End Computing (ICHEC) (project tcphy075c) and the Trinity Centre for High Performance
Computing (TCHPC) for use of computational resources.

\bibliography{library}

\begin{thebibliography}{21}%
\makeatletter
\providecommand \@ifxundefined [1]{%
 \@ifx{#1\undefined}
}%
\providecommand \@ifnum [1]{%
 \ifnum #1\expandafter \@firstoftwo
 \else \expandafter \@secondoftwo
 \fi
}%
\providecommand \@ifx [1]{%
 \ifx #1\expandafter \@firstoftwo
 \else \expandafter \@secondoftwo
 \fi
}%
\providecommand \natexlab [1]{#1}%
\providecommand \enquote  [1]{``#1''}%
\providecommand \bibnamefont  [1]{#1}%
\providecommand \bibfnamefont [1]{#1}%
\providecommand \citenamefont [1]{#1}%
\providecommand \href@noop [0]{\@secondoftwo}%
\providecommand \href [0]{\begingroup \@sanitize@url \@href}%
\providecommand \@href[1]{\@@startlink{#1}\@@href}%
\providecommand \@@href[1]{\endgroup#1\@@endlink}%
\providecommand \@sanitize@url [0]{\catcode `\\12\catcode `\$12\catcode
  `\&12\catcode `\#12\catcode `\^12\catcode `\_12\catcode `\%12\relax}%
\providecommand \@@startlink[1]{}%
\providecommand \@@endlink[0]{}%
\providecommand \url  [0]{\begingroup\@sanitize@url \@url }%
\providecommand \@url [1]{\endgroup\@href {#1}{\urlprefix }}%
\providecommand \urlprefix  [0]{URL }%
\providecommand \Eprint [0]{\href }%
\providecommand \doibase [0]{http://dx.doi.org/}%
\providecommand \selectlanguage [0]{\@gobble}%
\providecommand \bibinfo  [0]{\@secondoftwo}%
\providecommand \bibfield  [0]{\@secondoftwo}%
\providecommand \translation [1]{[#1]}%
\providecommand \BibitemOpen [0]{}%
\providecommand \bibitemStop [0]{}%
\providecommand \bibitemNoStop [0]{.\EOS\space}%
\providecommand \EOS [0]{\spacefactor3000\relax}%
\providecommand \BibitemShut  [1]{\csname bibitem#1\endcsname}%
\let\auto@bib@innerbib\@empty
\bibitem [{\citenamefont {Brataas}\ \emph {et~al.}(2012)\citenamefont
  {Brataas}, \citenamefont {Kent},\ and\ \citenamefont {Ohno}}]{Brataas2012}%
  \BibitemOpen
  \bibfield  {author} {\bibinfo {author} {\bibfnamefont {A.}~\bibnamefont
  {Brataas}}, \bibinfo {author} {\bibfnamefont {A.~D.}\ \bibnamefont {Kent}}, \
  and\ \bibinfo {author} {\bibfnamefont {H.}~\bibnamefont {Ohno}},\ }\href
  {\doibase 10.1038/nmat3311} {\bibfield  {journal} {\bibinfo  {journal}
  {Nature Materials}\ }\textbf {\bibinfo {volume} {11}},\ \bibinfo {pages}
  {372} (\bibinfo {year} {2012})}\BibitemShut {NoStop}%
\bibitem [{\citenamefont {Ralph}\ and\ \citenamefont
  {Stiles}(2008)}]{Ralph2008}%
  \BibitemOpen
  \bibfield  {author} {\bibinfo {author} {\bibfnamefont {D.~C.}\ \bibnamefont
  {Ralph}}\ and\ \bibinfo {author} {\bibfnamefont {M.~D.}\ \bibnamefont
  {Stiles}},\ }\href {\doibase 10.1016/j.jmmm.2007.12.019} {\bibfield
  {journal} {\bibinfo  {journal} {Journal of Magnetism and Magnetic Materials}\
  }\textbf {\bibinfo {volume} {320}},\ \bibinfo {pages} {1190} (\bibinfo {year}
  {2008})},\ \Eprint {http://arxiv.org/abs/0711.4608} {arXiv:0711.4608}
  \BibitemShut {NoStop}%
\bibitem [{\citenamefont {Slonczewski}(1996)}]{Slonczewski1996}%
  \BibitemOpen
  \bibfield  {author} {\bibinfo {author} {\bibfnamefont {J.}~\bibnamefont
  {Slonczewski}},\ }\href {\doibase 10.1016/0304-8853(96)00062-5} {\bibfield
  {journal} {\bibinfo  {journal} {Journal of Magnetism and Magnetic Materials}\
  }\textbf {\bibinfo {volume} {159}},\ \bibinfo {pages} {L1} (\bibinfo {year}
  {1996})}\BibitemShut {NoStop}%
\bibitem [{\citenamefont {Stamps}\ \emph {et~al.}(2014)\citenamefont {Stamps},
  \citenamefont {Breitkreutz}, \citenamefont {{\AA}kerman}, \citenamefont
  {Chumak}, \citenamefont {Otani}, \citenamefont {Bauer}, \citenamefont
  {Thiele}, \citenamefont {Bowen}, \citenamefont {Majetich}, \citenamefont
  {Kl{\"{a}}ui}, \citenamefont {Prejbeanu}, \citenamefont {Dieny},
  \citenamefont {Dempsey},\ and\ \citenamefont {Hillebrands}}]{Stamps2014}%
  \BibitemOpen
  \bibfield  {author} {\bibinfo {author} {\bibfnamefont {R.~L.}\ \bibnamefont
  {Stamps}}, \bibinfo {author} {\bibfnamefont {S.}~\bibnamefont {Breitkreutz}},
  \bibinfo {author} {\bibfnamefont {J.}~\bibnamefont {{\AA}kerman}}, \bibinfo
  {author} {\bibfnamefont {A.~V.}\ \bibnamefont {Chumak}}, \bibinfo {author}
  {\bibfnamefont {Y.}~\bibnamefont {Otani}}, \bibinfo {author} {\bibfnamefont
  {G.~E.~W.}\ \bibnamefont {Bauer}}, \bibinfo {author} {\bibfnamefont {J.-U.}\
  \bibnamefont {Thiele}}, \bibinfo {author} {\bibfnamefont {M.}~\bibnamefont
  {Bowen}}, \bibinfo {author} {\bibfnamefont {S.~a.}\ \bibnamefont {Majetich}},
  \bibinfo {author} {\bibfnamefont {M.}~\bibnamefont {Kl{\"{a}}ui}}, \bibinfo
  {author} {\bibfnamefont {I.~L.}\ \bibnamefont {Prejbeanu}}, \bibinfo {author}
  {\bibfnamefont {B.}~\bibnamefont {Dieny}}, \bibinfo {author} {\bibfnamefont
  {N.~M.}\ \bibnamefont {Dempsey}}, \ and\ \bibinfo {author} {\bibfnamefont
  {B.}~\bibnamefont {Hillebrands}},\ }\href {\doibase
  10.1088/0022-3727/47/33/333001} {\bibfield  {journal} {\bibinfo  {journal}
  {Journal of Physics D: Applied Physics}\ }\textbf {\bibinfo {volume} {47}},\
  \bibinfo {pages} {333001} (\bibinfo {year} {2014})},\ \Eprint
  {http://arxiv.org/abs/1410.6404} {arXiv:1410.6404} \BibitemShut {NoStop}%
\bibitem [{\citenamefont {Heiliger}\ and\ \citenamefont
  {Stiles}(2008)}]{Heiliger2008}%
  \BibitemOpen
  \bibfield  {author} {\bibinfo {author} {\bibfnamefont {C.}~\bibnamefont
  {Heiliger}}\ and\ \bibinfo {author} {\bibfnamefont {M.~D.}\ \bibnamefont
  {Stiles}},\ }\href {\doibase 10.1103/PhysRevLett.100.186805} {\bibfield
  {journal} {\bibinfo  {journal} {Physical Review Letters}\ }\textbf {\bibinfo
  {volume} {100}},\ \bibinfo {pages} {186805} (\bibinfo {year}
  {2008})}\BibitemShut {NoStop}%
\bibitem [{\citenamefont {Haney}\ \emph
  {et~al.}(2007{\natexlab{a}})\citenamefont {Haney}, \citenamefont {Waldron},
  \citenamefont {Duine}, \citenamefont {N{\'{u}}{\~{n}}ez}, \citenamefont
  {Guo},\ and\ \citenamefont {MacDonald}}]{Haney2008}%
  \BibitemOpen
  \bibfield  {author} {\bibinfo {author} {\bibfnamefont {P.~M.}\ \bibnamefont
  {Haney}}, \bibinfo {author} {\bibfnamefont {D.}~\bibnamefont {Waldron}},
  \bibinfo {author} {\bibfnamefont {R.~A.}\ \bibnamefont {Duine}}, \bibinfo
  {author} {\bibfnamefont {A.~S.}\ \bibnamefont {N{\'{u}}{\~{n}}ez}}, \bibinfo
  {author} {\bibfnamefont {H.}~\bibnamefont {Guo}}, \ and\ \bibinfo {author}
  {\bibfnamefont {A.~H.}\ \bibnamefont {MacDonald}},\ }\href {\doibase
  10.1103/PhysRevB.77.059901} {\bibfield  {journal} {\bibinfo  {journal}
  {Physical Review B}\ }\textbf {\bibinfo {volume} {76}},\ \bibinfo {pages}
  {024404} (\bibinfo {year} {2007}{\natexlab{a}})}\BibitemShut {NoStop}%
\bibitem [{\citenamefont {Berkov}\ and\ \citenamefont
  {Miltat}(2008)}]{Berkov2008}%
  \BibitemOpen
  \bibfield  {author} {\bibinfo {author} {\bibfnamefont {D.~V.}\ \bibnamefont
  {Berkov}}\ and\ \bibinfo {author} {\bibfnamefont {J.}~\bibnamefont
  {Miltat}},\ }\href {\doibase 10.1016/j.jmmm.2007.12.023} {\bibfield
  {journal} {\bibinfo  {journal} {Journal of Magnetism and Magnetic Materials}\
  }\textbf {\bibinfo {volume} {320}},\ \bibinfo {pages} {1238} (\bibinfo {year}
  {2008})},\ \Eprint {http://arxiv.org/abs/0710.5924} {arXiv:0710.5924}
  \BibitemShut {NoStop}%
\bibitem [{\citenamefont {Chureemart}\ \emph {et~al.}(2017)\citenamefont
  {Chureemart}, \citenamefont {Cuadrado}, \citenamefont {Chureemart},\ and\
  \citenamefont {Chantrell}}]{Chureemart2017}%
  \BibitemOpen
  \bibfield  {author} {\bibinfo {author} {\bibfnamefont {J.}~\bibnamefont
  {Chureemart}}, \bibinfo {author} {\bibfnamefont {R.}~\bibnamefont
  {Cuadrado}}, \bibinfo {author} {\bibfnamefont {P.}~\bibnamefont
  {Chureemart}}, \ and\ \bibinfo {author} {\bibfnamefont {R.}~\bibnamefont
  {Chantrell}},\ }\href {\doibase 10.1016/j.jmmm.2017.07.085} {\bibfield
  {journal} {\bibinfo  {journal} {Journal of Magnetism and Magnetic Materials}\
  }\textbf {\bibinfo {volume} {443}},\ \bibinfo {pages} {287} (\bibinfo {year}
  {2017})}\BibitemShut {NoStop}%
\bibitem [{\citenamefont {Ellis}\ \emph {et~al.}(2015)\citenamefont {Ellis},
  \citenamefont {Evans}, \citenamefont {Ostler}, \citenamefont {Barker},
  \citenamefont {Atxitia}, \citenamefont {Chubykalo-Fesenko},\ and\
  \citenamefont {Chantrell}}]{Ellis2015}%
  \BibitemOpen
  \bibfield  {author} {\bibinfo {author} {\bibfnamefont {M.~O.~A.}\
  \bibnamefont {Ellis}}, \bibinfo {author} {\bibfnamefont {R.~F.~L.}\
  \bibnamefont {Evans}}, \bibinfo {author} {\bibfnamefont {T.~A.}\ \bibnamefont
  {Ostler}}, \bibinfo {author} {\bibfnamefont {J.}~\bibnamefont {Barker}},
  \bibinfo {author} {\bibfnamefont {U.}~\bibnamefont {Atxitia}}, \bibinfo
  {author} {\bibfnamefont {O.}~\bibnamefont {Chubykalo-Fesenko}}, \ and\
  \bibinfo {author} {\bibfnamefont {R.~W.}\ \bibnamefont {Chantrell}},\ }\href
  {\doibase 10.1063/1.4930971} {\bibfield  {journal} {\bibinfo  {journal} {Low
  Temperature Physics}\ }\textbf {\bibinfo {volume} {41}},\ \bibinfo {pages}
  {705} (\bibinfo {year} {2015})}\BibitemShut {NoStop}%
\bibitem [{\citenamefont {Evans}\ \emph {et~al.}(2014)\citenamefont {Evans},
  \citenamefont {Fan}, \citenamefont {Chureemart}, \citenamefont {Ostler},
  \citenamefont {Ellis},\ and\ \citenamefont {Chantrell}}]{Evans2014}%
  \BibitemOpen
  \bibfield  {author} {\bibinfo {author} {\bibfnamefont {R.~F.~L.}\
  \bibnamefont {Evans}}, \bibinfo {author} {\bibfnamefont {W.~J.}\ \bibnamefont
  {Fan}}, \bibinfo {author} {\bibfnamefont {P.}~\bibnamefont {Chureemart}},
  \bibinfo {author} {\bibfnamefont {T.~A.}\ \bibnamefont {Ostler}}, \bibinfo
  {author} {\bibfnamefont {M.~O.~A.}\ \bibnamefont {Ellis}}, \ and\ \bibinfo
  {author} {\bibfnamefont {R.~W.}\ \bibnamefont {Chantrell}},\ }\href {\doibase
  10.1088/0953-8984/26/10/103202} {\bibfield  {journal} {\bibinfo  {journal}
  {Journal of Physics: Condensed Matter}\ }\textbf {\bibinfo {volume} {26}},\
  \bibinfo {pages} {103202} (\bibinfo {year} {2014})}\BibitemShut {NoStop}%
\bibitem [{\citenamefont {Rocha}\ \emph {et~al.}(2006)\citenamefont {Rocha},
  \citenamefont {Garc{\'{i}}a-Su{\'{a}}rez}, \citenamefont {Bailey},
  \citenamefont {Lambert}, \citenamefont {Ferrer},\ and\ \citenamefont
  {Sanvito}}]{Rocha2006}%
  \BibitemOpen
  \bibfield  {author} {\bibinfo {author} {\bibfnamefont {A.~R.}\ \bibnamefont
  {Rocha}}, \bibinfo {author} {\bibfnamefont {V.~M.}\ \bibnamefont
  {Garc{\'{i}}a-Su{\'{a}}rez}}, \bibinfo {author} {\bibfnamefont
  {S.}~\bibnamefont {Bailey}}, \bibinfo {author} {\bibfnamefont
  {C.}~\bibnamefont {Lambert}}, \bibinfo {author} {\bibfnamefont
  {J.}~\bibnamefont {Ferrer}}, \ and\ \bibinfo {author} {\bibfnamefont
  {S.}~\bibnamefont {Sanvito}},\ }\href {\doibase 10.1103/PhysRevB.73.085414}
  {\bibfield  {journal} {\bibinfo  {journal} {Physical Review B}\ }\textbf
  {\bibinfo {volume} {73}},\ \bibinfo {pages} {085414} (\bibinfo {year}
  {2006})}\BibitemShut {NoStop}%
\bibitem [{\citenamefont {Rungger}\ \emph {et~al.}(2009)\citenamefont
  {Rungger}, \citenamefont {Mryasov},\ and\ \citenamefont
  {Sanvito}}]{Rungger2009}%
  \BibitemOpen
  \bibfield  {author} {\bibinfo {author} {\bibfnamefont {I.}~\bibnamefont
  {Rungger}}, \bibinfo {author} {\bibfnamefont {O.}~\bibnamefont {Mryasov}}, \
  and\ \bibinfo {author} {\bibfnamefont {S.}~\bibnamefont {Sanvito}},\ }\href
  {\doibase 10.1103/PhysRevB.79.094414} {\bibfield  {journal} {\bibinfo
  {journal} {Physical Review B}\ }\textbf {\bibinfo {volume} {79}},\ \bibinfo
  {pages} {094414} (\bibinfo {year} {2009})},\ \Eprint
  {http://arxiv.org/abs/0808.0902} {arXiv:0808.0902} \BibitemShut {NoStop}%
\bibitem [{\citenamefont {Soler}\ \emph {et~al.}(2002)\citenamefont {Soler},
  \citenamefont {Artacho}, \citenamefont {Gale}, \citenamefont {Garcia},
  \citenamefont {Junquera}, \citenamefont {Ordejon},\ and\ \citenamefont
  {Sanchez-Portal}}]{Soler2002}%
  \BibitemOpen
  \bibfield  {author} {\bibinfo {author} {\bibfnamefont {J.~M.}\ \bibnamefont
  {Soler}}, \bibinfo {author} {\bibfnamefont {E.}~\bibnamefont {Artacho}},
  \bibinfo {author} {\bibfnamefont {J.~D.}\ \bibnamefont {Gale}}, \bibinfo
  {author} {\bibfnamefont {A.}~\bibnamefont {Garcia}}, \bibinfo {author}
  {\bibfnamefont {J.}~\bibnamefont {Junquera}}, \bibinfo {author}
  {\bibfnamefont {P.}~\bibnamefont {Ordejon}}, \ and\ \bibinfo {author}
  {\bibfnamefont {D.}~\bibnamefont {Sanchez-Portal}},\ }\href {\doibase
  10.1088/0953-8984/14/11/302} {\bibfield  {journal} {\bibinfo  {journal}
  {Journal of Physics: Condensed Matter}\ }\textbf {\bibinfo {volume} {2745}},\
  \bibinfo {pages} {22} (\bibinfo {year} {2002})},\ \Eprint
  {http://arxiv.org/abs/0111138} {arXiv:0111138 [cond-mat]} \BibitemShut
  {NoStop}%
\bibitem [{\citenamefont {Haney}\ \emph
  {et~al.}(2007{\natexlab{b}})\citenamefont {Haney}, \citenamefont {Waldron},
  \citenamefont {Duine}, \citenamefont {N{\'{u}}{\~{n}}ez}, \citenamefont
  {Guo},\ and\ \citenamefont {MacDonald}}]{Haney2007}%
  \BibitemOpen
  \bibfield  {author} {\bibinfo {author} {\bibfnamefont {P.~M.}\ \bibnamefont
  {Haney}}, \bibinfo {author} {\bibfnamefont {D.}~\bibnamefont {Waldron}},
  \bibinfo {author} {\bibfnamefont {R.~A.}\ \bibnamefont {Duine}}, \bibinfo
  {author} {\bibfnamefont {A.~S.}\ \bibnamefont {N{\'{u}}{\~{n}}ez}}, \bibinfo
  {author} {\bibfnamefont {H.}~\bibnamefont {Guo}}, \ and\ \bibinfo {author}
  {\bibfnamefont {A.~H.}\ \bibnamefont {MacDonald}},\ }\href {\doibase
  10.1103/PhysRevB.75.174428} {\bibfield  {journal} {\bibinfo  {journal}
  {Physical Review B - Condensed Matter and Materials Physics}\ }\textbf
  {\bibinfo {volume} {75}},\ \bibinfo {pages} {174428} (\bibinfo {year}
  {2007}{\natexlab{b}})},\ \Eprint {http://arxiv.org/abs/0611599}
  {arXiv:0611599 [cond-mat]} \BibitemShut {NoStop}%
\bibitem [{\citenamefont {Stamenova}\ \emph {et~al.}(2016)\citenamefont
  {Stamenova}, \citenamefont {Mohebbi}, \citenamefont {Seyedyazdi},
  \citenamefont {Rungger},\ and\ \citenamefont {Sanvito}}]{Stamenova2016}%
  \BibitemOpen
  \bibfield  {author} {\bibinfo {author} {\bibfnamefont {M.}~\bibnamefont
  {Stamenova}}, \bibinfo {author} {\bibfnamefont {R.}~\bibnamefont {Mohebbi}},
  \bibinfo {author} {\bibfnamefont {J.}~\bibnamefont {Seyedyazdi}}, \bibinfo
  {author} {\bibfnamefont {I.}~\bibnamefont {Rungger}}, \ and\ \bibinfo
  {author} {\bibfnamefont {S.}~\bibnamefont {Sanvito}},\ }\href {\doibase
  10.1103/PhysRevB.95.060403} {\ \textbf {\bibinfo {volume} {060403}},\
  \bibinfo {pages} {1} (\bibinfo {year} {2016})},\ \Eprint
  {http://arxiv.org/abs/1611.07445} {arXiv:1611.07445} \BibitemShut {NoStop}%
\bibitem [{\citenamefont {Slonczewski}(2005)}]{Slonczewski2005}%
  \BibitemOpen
  \bibfield  {author} {\bibinfo {author} {\bibfnamefont {J.~C.}\ \bibnamefont
  {Slonczewski}},\ }\href {\doibase 10.1103/PhysRevB.71.024411} {\bibfield
  {journal} {\bibinfo  {journal} {Phys. Rev. B - Condens. Matter Mater. Phys.}\
  }\textbf {\bibinfo {volume} {71}},\ \bibinfo {pages} {1} (\bibinfo {year}
  {2005})},\ \Eprint {http://arxiv.org/abs/0404210} {arXiv:0404210 [cond-mat]}
  \BibitemShut {NoStop}%
\bibitem [{\citenamefont {Slonczewski}\ and\ \citenamefont
  {Sun}(2007)}]{Slonczewski2007}%
  \BibitemOpen
  \bibfield  {author} {\bibinfo {author} {\bibfnamefont {J.~C.}\ \bibnamefont
  {Slonczewski}}\ and\ \bibinfo {author} {\bibfnamefont {J.~Z.}\ \bibnamefont
  {Sun}},\ }\href {\doibase 10.1016/j.jmmm.2006.10.507} {\bibfield  {journal}
  {\bibinfo  {journal} {Journal of Magnetism and Magnetic Materials}\ }\textbf
  {\bibinfo {volume} {310}},\ \bibinfo {pages} {169} (\bibinfo {year}
  {2007})}\BibitemShut {NoStop}%
\bibitem [{\citenamefont {Xie}\ \emph {et~al.}(2016)\citenamefont {Xie},
  \citenamefont {Rungger}, \citenamefont {Munira}, \citenamefont {Stamenova},
  \citenamefont {Sanvito},\ and\ \citenamefont {Ghosh}}]{Xie2016}%
  \BibitemOpen
  \bibfield  {author} {\bibinfo {author} {\bibfnamefont {Y.}~\bibnamefont
  {Xie}}, \bibinfo {author} {\bibfnamefont {I.}~\bibnamefont {Rungger}},
  \bibinfo {author} {\bibfnamefont {K.}~\bibnamefont {Munira}}, \bibinfo
  {author} {\bibfnamefont {M.}~\bibnamefont {Stamenova}}, \bibinfo {author}
  {\bibfnamefont {S.}~\bibnamefont {Sanvito}}, \ and\ \bibinfo {author}
  {\bibfnamefont {A.~W.}\ \bibnamefont {Ghosh}},\ }\href {\doibase
  10.1002/9781118869239.ch4} {\bibfield  {journal} {\bibinfo  {journal}
  {Nanomagnetic Devices and Phenomena for Energy-Efficient Computing}\ ,\
  \bibinfo {pages} {91}} (\bibinfo {year} {2016})}\BibitemShut {NoStop}%
\bibitem [{\citenamefont {Hallal}\ \emph {et~al.}(2013)\citenamefont {Hallal},
  \citenamefont {Yang}, \citenamefont {Dieny},\ and\ \citenamefont
  {Chshiev}}]{Hallal2013}%
  \BibitemOpen
  \bibfield  {author} {\bibinfo {author} {\bibfnamefont {A.}~\bibnamefont
  {Hallal}}, \bibinfo {author} {\bibfnamefont {H.~X.}\ \bibnamefont {Yang}},
  \bibinfo {author} {\bibfnamefont {B.}~\bibnamefont {Dieny}}, \ and\ \bibinfo
  {author} {\bibfnamefont {M.}~\bibnamefont {Chshiev}},\ }\href {\doibase
  10.1103/PhysRevB.88.184423} {\bibfield  {journal} {\bibinfo  {journal}
  {Physical Review B}\ }\textbf {\bibinfo {volume} {88}},\ \bibinfo {pages}
  {184423} (\bibinfo {year} {2013})}\BibitemShut {NoStop}%
\bibitem [{\citenamefont {Iihama}\ \emph {et~al.}(2015)\citenamefont {Iihama},
  \citenamefont {Sasaki}, \citenamefont {Naganuma},\ and\ \citenamefont
  {Oogane}}]{Iihama2015}%
  \BibitemOpen
  \bibfield  {author} {\bibinfo {author} {\bibfnamefont {S.}~\bibnamefont
  {Iihama}}, \bibinfo {author} {\bibfnamefont {Y.}~\bibnamefont {Sasaki}},
  \bibinfo {author} {\bibfnamefont {H.}~\bibnamefont {Naganuma}}, \ and\
  \bibinfo {author} {\bibfnamefont {M.}~\bibnamefont {Oogane}},\ }\href
  {\doibase 10.1088/0022-3727/49/3/035002} {\bibfield  {journal} {\bibinfo
  {journal} {Journal of Physics D: Applied Physics}\ }\textbf {\bibinfo
  {volume} {49}},\ \bibinfo {pages} {35002} (\bibinfo {year}
  {2015})}\BibitemShut {NoStop}%
\bibitem [{\citenamefont {Callen}\ and\ \citenamefont
  {Callen}(1963)}]{Callen1963}%
  \BibitemOpen
  \bibfield  {author} {\bibinfo {author} {\bibfnamefont {E.~R.}\ \bibnamefont
  {Callen}}\ and\ \bibinfo {author} {\bibfnamefont {H.~B.}\ \bibnamefont
  {Callen}},\ }\href {\doibase 10.1103/PhysRev.129.578} {\bibfield  {journal}
  {\bibinfo  {journal} {Physical Review}\ }\textbf {\bibinfo {volume} {129}},\
  \bibinfo {pages} {578} (\bibinfo {year} {1963})}\BibitemShut {NoStop}%
\end{thebibliography}%

\end{document}